# Dimer in the Hubbard model. Exact and approximate solutions.

Gennadiy Ivanovich Mironov*[a]

[a] Head of the Department of Physics and Materials Science,
Doctor of Physical and Mathematical Sciences, Prof. G.I. Mironov
Mari State University
Kremlevskaya Street 44, Yoshkar-Ola (Russia)
E-mail: mirgi@marsu.ru

Within the tight-binding model taking into account the Coulomb repulsion of electrons at one site (the Hubbard model), an exact calculation of the Fourier transform of the anticommutator Green's function of the $C_2$ dimer as a structural element of fullerene, the poles of which determine the energy spectrum of the studied nanosystem, was performed. Graphic representations of the equation for the chemical potential and the density of state of electrons were obtained. A solution for the dimer was obtained within the approximation of static fluctuations. The Fourier transform of the anticommutator Green's function, obtained in the approximation of static fluctuations, coincided with a similar function obtained within the exact solution. The ionization energy and electron affinity were calculated and found to be 11.87 eV and 3.39 eV, respectively. It was concluded that the method for solving the Hubbard model for the $C_2$ dimer in the approximation of static fluctuations proposed in the work can be a fairly adequate method for theoretical study of strongly correlated systems.

**Introduction**

The Hubbard model was proposed to explain the ferromagnetism of collectivized electrons and to describe the metal-insulator transition in transition metals more than sixty years ago [1, 2]. Independently of [1, 2], a similar model was proposed in [3, 4]. Moreover, the model proposed in [1-4] was a special case of the polar model of metal proposed by Shubin and Vonsovsky almost thirty years earlier [5]. Interest in the Hubbard model increased significantly after the discovery of high-temperature superconductivity in copper oxide compounds [6, 7], since in the search for high-temperature superconductivity mechanisms, at the initial stage, special hopes were placed on various correlation mechanisms that exist in the Hubbard model and related models [8-11].

Subsequently, interest in the Hubbard model was fueled by the fact that the system of π-electrons in carbon fullerenes [12], nanotubes [13], graphene [14, 15], and biphenylene [16] behaved as a strongly correlated system. Indeed, in each of the considered cases of new allotropic states of carbon (fullerene, nanotube, graphene, biphenylene), each carbon atom is surrounded by three nearest neighboring carbon atoms. S-p hybridization in the case of these states leads to the fact that the three σ-electrons of each carbon atom form three covalent bonds with three neighboring atoms. The energy level of the fourth π-electron is located above the energy level of the σ-electrons, and the wave function of the π-electron overlaps with the wave functions of all three neighboring π-electrons. For this reason, the π-electron can move to any of the three neighboring atoms. The π-electrons are responsible for the electronic and transport properties of the carbon nanosystems under consideration; they make the main contribution to these properties. If a π-electron has moved to a neighboring atom in accordance with the Pauli principle, then two π-electrons located at one site will experience strong Coulomb repulsion from each other; the energy of their Coulomb repulsion will be significantly greater than the transfer integral of the π-electron from a site to a neighboring site of the nanosystem. Therefore, a system of strongly correlated π-electrons will arise in the carbon nanosystems under consideration, which is responsible for the main properties of carbon nanosystems. Since π-electrons are responsible for the main properties in carbon nanosystems, the influence of σ-electrons on the properties of carbon nanosystems can be neglected; the behavior of carbon nanosystems can be modeled within the framework of the single-band Hubbard Hamiltonian.

The exact solution of the Hubbard model is known only for the one-dimensional model [17]. The most interesting feature of the solution in [17] is that the energy gap $\Delta$ is nonzero for any values of the Coulomb potential U, only for U→0 the gap width $\Delta$→0. Thus, the one-dimensional system remains dielectric for any interaction. Obviously, this is due to the peculiarity of the one-dimensional case; it is known that a one-dimensional system is always unstable with respect to period doubling, at which a gap appears at the Fermi level [18]. In this case, period doubling and the presence of a gap are apparently associated with the emergence of antiferromagnetism. Analysis of the spin structure in the one-dimensional Hubbard model in [17] was not carried out [19]. The exact solution shows that the one-dimensional Hubbard model has an antiferromagnetic dielectric ground state. There is no Mott transition [20] in the parameter U in the case of a one-dimensional system. Excited states in the one-dimensional Hubbard model were studied in [21], the thermodynamics of the one-dimensional Hubbard model was studied in [22]. In [23], the one-dimensional Hubbard model was solved in the presence of a magnetic field. When solving one-dimensional models, the Bethe ansatz was used [24]. These solutions of the one-dimensional model were of a fundamental nature, they allowed us to control the adequacy of approximate methods for solving the two-dimensional and three-dimensional Hubbard models. Of the many approximate methods, we will note only the following. In [25], the half-filled Hubbard model at zero temperature was investigated using the renormalization group method in real space. Although the method is approximate, a good qualitative and quantitative description of the systems for which the Hubbard Hamiltonian was proposed was obtained in [25]. In the one-dimensional case, the ground state of the system was shown to be dielectric, and the calculation of the ground state energy showed good agreement with the exact results of [17–23]. Note that a problem similar to [25] was previously solved by the same renormalization group method in [26], where an expression for the ground state energy of the one-dimensional Hubbard model was also obtained for comparison with the exact solution. Comparison of the results of [25, 26] with [17] showed that the energy obtained by the approximate methods lies higher than the exactly calculated one (see Fig. 5 in [25] and Fig. 1 in [26]). The solution [25] is closer to the exact solution than that calculated in [26], the authors of [25] explained this by the fact that at each iteration step they took into account only the underlying energy states. For comparison with the existing exact result [17], the author of [27] considered a special case of the one-dimensional Hubbard model in the static fluctuation approximation. Comparison of the results showed that in the cases of strong and weak couplings, the ground state energies in the case of the exact solution of the one-dimensional model and in the case of the static fluctuation approximation practically coincided, and in the case of intermediate correlations there was good quantitative agreement. The difference was that the graph for the ground state energy was slightly higher compared to the graph obtained from the exact solution. In this region, there was a slight overestimation of the role of the Coulomb interaction. If in works [25, 26] in cases when the Coulomb potential is equal to zero and when the Coulomb potential tends to infinity, the energies of the ground state differed from the energies of the ground state in the exact solution, then in [27] these values coincided. If in the case of weak coupling, when the ratio of the Coulomb potential U to the transfer integral B is equal to two, the discrepancy with the exact solution is 16% and 9% in works [26] and [25], respectively, (see Table 3 in [25]), then in the case of [27] the discrepancy is 2%.

We note one more point in view of the obvious importance of carbon dimers existing in the microstructure of steel [28, 29] and their possible interactions with other atoms – it is important to develop effective quantum-mechanical and quantum-field methods for solving the dimer. From this point of view, the present work acquires a certain significance.

**Exact solution of the dimer as a strongly correlated system in the Hubbard model**

The Hamiltonian of a dimer in the Hubbard model [1, 2] has the form:

$$\hat{H} = \varepsilon \sum_{f=1}^{2} (\hat{n}_{f\uparrow} + \hat{n}_{f\downarrow}) + B \sum_{f \neq f'} (a_{f\uparrow}^{+} a_{f'\uparrow} + a_{f\downarrow}^{+} a_{f'\downarrow}) + U \sum_{f=1}^{2} \hat{n}_{f\uparrow} \hat{n}_{f\downarrow}. \qquad (1)$$

The first term in (1) describes the self-energy of π-electrons, the second term is the energy of electron transfer from a site to a neighboring site of the dimer, and the third term is the energy of Coulomb repulsion of two π-electrons that end up on the same site of the dimer. $\hat{n}_{f\uparrow}$ is the operator of the number of π-electrons at the site f of the dimer with the spin projection ↑, $a_{f\uparrow}^{+}$ is the operator of π-electron creation, and $a_{f'\uparrow}$ is the operator of π-electron annihilation.

By writing the π-electron creation operator in the Heisenberg representation, we can obtain the following 22 equations of motion for the second quantization operators ($i$ in the equations takes the values 1 and 2, $j$, respectively, takes the values 2 and 1, $i \neq j$, besides $\tau = it$ is imaginary time):

$$\frac{da_{i\uparrow}^+}{d\tau} = \varepsilon a_{i\uparrow}^+ + B a_{j\uparrow}^+ + U \hat{n}_{i\downarrow} a_{i\uparrow}^+, \tag{2}$$

$$\frac{d\hat{n}_{i\downarrow} a_{i\uparrow}^+}{d\tau} = (\varepsilon + U)\hat{n}_{i\downarrow} a_{i\uparrow}^+ + B\left(a_{j\downarrow}^+ a_{i\downarrow} a_{i\uparrow}^+ - a_{i\downarrow}^+ a_{j\downarrow} a_{i\uparrow}^+ + \hat{n}_{i\downarrow} a_{j\uparrow}^+\right), \tag{3}$$

$$\frac{d a_{j\downarrow}^+ a_{i\downarrow} a_{i\uparrow}^+}{d\tau} = \varepsilon\, a_{j\downarrow}^+ a_{i\downarrow} a_{i\uparrow}^+ + B\left(\hat{n}_{i\downarrow} a_{i\uparrow}^+ - \hat{n}_{j\downarrow} a_{i\uparrow}^+ + a_{j\downarrow}^+ a_{i\downarrow} a_{j\uparrow}^+\right) + U \hat{n}_{j\uparrow} a_{j\downarrow}^+ a_{i\downarrow} a_{i\uparrow}^+, \tag{4}$$

$$\frac{d a_{i\downarrow}^+ a_{j\downarrow} a_{i\uparrow}^+}{d\tau} = (\varepsilon + U)\, a_{i\downarrow}^+ a_{j\downarrow} a_{i\uparrow}^+ + B\left(\hat{n}_{j\downarrow} a_{i\uparrow}^+ - \hat{n}_{i\downarrow} a_{i\uparrow}^+ + a_{i\downarrow}^+ a_{j\downarrow} a_{j\uparrow}^+\right) - U \hat{n}_{j\uparrow} a_{i\downarrow}^+ a_{j\downarrow} a_{i\uparrow}^+, \tag{5}$$

$$\frac{d\hat{n}_{i\downarrow} a_{j\uparrow}^+}{d\tau} = \varepsilon\, \hat{n}_{i\downarrow} a_{j\uparrow}^+ + B\left(a_{j\downarrow}^+ a_{i\downarrow} a_{j\uparrow}^+ - a_{i\downarrow}^+ a_{j\downarrow} a_{j\uparrow}^+ + \hat{n}_{i\downarrow} a_{i\uparrow}^+\right) + U \hat{n}_{i\downarrow} \hat{n}_{j\downarrow} a_{j\uparrow}^+, \tag{6}$$

$$\frac{d\hat{n}_{i\downarrow}\hat{n}_{j\downarrow} a_{j\uparrow}^+}{d\tau} = (\varepsilon + U)\hat{n}_{i\downarrow}\hat{n}_{j\downarrow} a_{j\uparrow}^+ + B\hat{n}_{i\downarrow}\hat{n}_{j\downarrow} a_{i\uparrow}^+, \tag{7}$$

$$\frac{d\hat{n}_{j\uparrow} a_{j\downarrow}^+ a_{i\downarrow} a_{i\uparrow}^+}{d\tau} = (\varepsilon + U)\hat{n}_{j\uparrow} a_{j\downarrow}^+ a_{i\downarrow} a_{i\uparrow}^+ + B\left(\hat{n}_{i\uparrow} a_{j\downarrow}^+ a_{i\downarrow} a_{j\uparrow}^+ - \hat{n}_{j\uparrow}\hat{n}_{j\downarrow} a_{i\uparrow}^+ + \hat{n}_{j\uparrow}\hat{n}_{i\downarrow} a_{i\uparrow}^+\right), \tag{8}$$

$$\frac{d\hat{n}_{j\uparrow} a_{i\downarrow}^+ a_{j\downarrow} a_{i\uparrow}^+}{d\tau} = \varepsilon\, \hat{n}_{j\uparrow} a_{i\downarrow}^+ a_{j\downarrow} a_{i\uparrow}^+ + B\left(\hat{n}_{i\uparrow} a_{i\downarrow}^+ a_{j\downarrow} a_{j\uparrow}^+ - \hat{n}_{j\uparrow}\hat{n}_{i\downarrow} a_{i\uparrow}^+ + \hat{n}_{j\uparrow}\hat{n}_{j\downarrow} a_{i\uparrow}^+\right), \tag{9}$$

$$\frac{d\hat{n}_{j\uparrow}\hat{n}_{i\downarrow} a_{i\uparrow}^+}{d\tau} = (\varepsilon + U)\hat{n}_{j\uparrow}\hat{n}_{i\downarrow} a_{i\uparrow}^+ + B\left(\hat{n}_{i\uparrow}\hat{n}_{i\downarrow} a_{j\uparrow}^+ - \hat{n}_{j\uparrow} a_{i\downarrow}^+ a_{j\downarrow} a_{i\uparrow}^+ + \hat{n}_{j\uparrow} a_{j\downarrow}^+ a_{i\downarrow} a_{i\uparrow}^+\right), \tag{10}$$

$$\frac{d\hat{n}_{j\uparrow}\hat{n}_{j\downarrow} a_{i\uparrow}^+}{d\tau} = (\varepsilon + U)\hat{n}_{j\uparrow}\hat{n}_{j\downarrow} a_{i\uparrow}^+ + B\left(\hat{n}_{i\uparrow}\hat{n}_{j\downarrow} a_{j\uparrow}^+ - \hat{n}_{j\uparrow} a_{j\downarrow}^+ a_{i\downarrow} a_{i\uparrow}^+ + \hat{n}_{j\uparrow} a_{i\downarrow}^+ a_{j\downarrow} a_{i\uparrow}^+\right)$$
$$+ U\hat{n}_{j\uparrow}\hat{n}_{j\downarrow}\hat{n}_{i\downarrow} a_{i\uparrow}^+, \tag{11}$$

$$\frac{d\hat{n}_{j\uparrow}\hat{n}_{j\downarrow}\hat{n}_{i\downarrow} a_{i\uparrow}^+}{d\tau} = (\varepsilon + U)\hat{n}_{j\uparrow}\hat{n}_{j\downarrow}\hat{n}_{i\downarrow} a_{i\uparrow}^+ + B\hat{n}_{i\uparrow}\hat{n}_{i\downarrow}\hat{n}_{j\downarrow} a_{j\uparrow}^+. \tag{12}$$

By solving this system of closed differential equations, we can obtain, for example, the following expression for the Fourier transform of the anticommutator Green's function in the ground state:

$$\langle\langle a_{1\uparrow}^+|a_{1\uparrow}\rangle\rangle_E$$

$$= \frac{i}{2\pi}\frac{1}{4}\left\{\frac{2 - 3\langle\hat{n}_{1\downarrow}\rangle - \langle a_{1\uparrow}^+ a_{2\uparrow}\rangle + \langle\hat{n}_{1\uparrow}\hat{n}_{2\uparrow}\rangle + \langle\hat{n}_{1\uparrow}a_{1\downarrow}^+ a_{2\downarrow}\rangle + \langle\hat{n}_{1\uparrow}a_{2\downarrow}^+ a_{1\downarrow}\rangle + \langle a_{2\uparrow}^+ a_{2\downarrow}^+ a_{1\uparrow}a_{2\downarrow}\rangle + \langle a_{2\uparrow}^+ a_{1\downarrow}^+ a_{1\uparrow}a_{2\downarrow}\rangle}{E - \varepsilon - B}\right.$$

$$+ \frac{2 - 3\langle\hat{n}_{1\downarrow}\rangle + \langle a_{1\uparrow}^+ a_{2\uparrow}\rangle + \langle\hat{n}_{1\uparrow}\hat{n}_{2\uparrow}\rangle - \langle\hat{n}_{1\uparrow}a_{1\downarrow}^+ a_{2\downarrow}\rangle - \langle\hat{n}_{1\uparrow}a_{2\downarrow}^+ a_{1\downarrow}\rangle + \langle a_{2\uparrow}^+ a_{2\downarrow}^+ a_{1\uparrow}a_{2\downarrow}\rangle + \langle a_{2\uparrow}^+ a_{1\downarrow}^+ a_{1\uparrow}a_{2\downarrow}\rangle}{E - \varepsilon + B}$$

$$+ \frac{\langle\hat{n}_{1\downarrow}\rangle - \langle a_{1\uparrow}^+ a_{2\uparrow}\rangle + \langle\hat{n}_{1\uparrow}\hat{n}_{2\uparrow}\rangle + \langle\hat{n}_{1\uparrow}a_{1\downarrow}^+ a_{2\downarrow}\rangle + \langle\hat{n}_{1\uparrow}a_{2\downarrow}^+ a_{1\downarrow}\rangle + \langle a_{2\uparrow}^+ a_{2\downarrow}^+ a_{1\uparrow}a_{2\downarrow}\rangle + \langle a_{2\uparrow}^+ a_{1\downarrow}^+ a_{1\uparrow}a_{2\downarrow}\rangle}{E - \varepsilon - U - B}$$

$$+ \frac{\langle\hat{n}_{1\downarrow}\rangle + \langle a_{1\uparrow}^+ a_{2\uparrow}\rangle + \langle\hat{n}_{1\uparrow}\hat{n}_{2\uparrow}\rangle - \langle\hat{n}_{1\uparrow}a_{1\downarrow}^+ a_{2\downarrow}\rangle - \langle\hat{n}_{1\uparrow}a_{2\downarrow}^+ a_{1\downarrow}\rangle + \langle a_{2\uparrow}^+ a_{2\downarrow}^+ a_{1\uparrow}a_{2\downarrow}\rangle + \langle a_{2\uparrow}^+ a_{1\downarrow}^+ a_{1\uparrow}a_{2\downarrow}\rangle}{E - \varepsilon - U + B}$$

$$+ \frac{\langle\hat{n}_{1\downarrow}\rangle + \langle a_{1\uparrow}^+ a_{2\uparrow}\rangle - \langle\hat{n}_{1\uparrow}\hat{n}_{2\uparrow}\rangle - \langle\hat{n}_{1\uparrow}a_{1\downarrow}^+ a_{2\downarrow}\rangle - \langle\hat{n}_{1\uparrow}a_{2\downarrow}^+ a_{1\downarrow}\rangle - \langle a_{2\uparrow}^+ a_{2\downarrow}^+ a_{1\uparrow}a_{2\downarrow}\rangle - \langle a_{2\uparrow}^+ a_{1\downarrow}^+ a_{1\uparrow}a_{2\downarrow}\rangle}{(E - \varepsilon - U/2 - \alpha/2 + B)/(1 + 4B/\alpha)} \quad (13)$$

$$+ \frac{\langle\hat{n}_{1\downarrow}\rangle + \langle a_{1\uparrow}^+ a_{2\uparrow}\rangle - \langle\hat{n}_{1\uparrow}\hat{n}_{2\uparrow}\rangle - \langle\hat{n}_{1\uparrow}a_{1\downarrow}^+ a_{2\downarrow}\rangle - \langle\hat{n}_{1\uparrow}a_{2\downarrow}^+ a_{1\downarrow}\rangle - \langle a_{2\uparrow}^+ a_{2\downarrow}^+ a_{1\uparrow}a_{2\downarrow}\rangle - \langle a_{2\uparrow}^+ a_{1\downarrow}^+ a_{1\uparrow}a_{2\downarrow}\rangle}{\left(E - \varepsilon - \frac{U}{2} + \alpha/2 + B\right)/(1 - 4B/\alpha)}$$

$$+ \frac{\langle\hat{n}_{1\downarrow}\rangle - \langle a_{1\uparrow}^+ a_{2\uparrow}\rangle - \langle\hat{n}_{1\uparrow}\hat{n}_{2\uparrow}\rangle + \langle\hat{n}_{1\uparrow}a_{1\downarrow}^+ a_{2\downarrow}\rangle + \langle\hat{n}_{1\uparrow}a_{2\downarrow}^+ a_{1\downarrow}\rangle - \langle a_{2\uparrow}^+ a_{2\downarrow}^+ a_{1\uparrow}a_{2\downarrow}\rangle - \langle a_{2\uparrow}^+ a_{1\downarrow}^+ a_{1\uparrow}a_{2\downarrow}\rangle}{\left(E - \varepsilon - \frac{U}{2} + \frac{\alpha}{2} - B\right)/(1 + 4B/\alpha)}$$

$$+ \left.\frac{\langle\hat{n}_{1\downarrow}\rangle - \langle a_{1\uparrow}^+ a_{2\uparrow}\rangle - \langle\hat{n}_{1\uparrow}\hat{n}_{2\uparrow}\rangle + \langle\hat{n}_{1\uparrow}a_{1\downarrow}^+ a_{2\downarrow}\rangle + \langle\hat{n}_{1\uparrow}a_{2\downarrow}^+ a_{1\downarrow}\rangle - \langle a_{2\uparrow}^+ a_{2\downarrow}^+ a_{1\uparrow}a_{2\downarrow}\rangle - \langle a_{2\uparrow}^+ a_{1\downarrow}^+ a_{1\uparrow}a_{2\downarrow}\rangle}{\left(E - \varepsilon - \frac{U}{2} - \frac{\alpha}{2} - B\right)/(1 - 4B/\alpha)}\right\}.$$

The constant value α has the form: $\alpha = \sqrt{U^2 + 16B^2}$.

Formula (13) is characterized by eight features, for example, in the case of the last fraction in (13), the pole occurs in the case of $E - \varepsilon - \frac{U}{2} - \frac{\alpha}{2} - B = 0$, the rest of this fraction, including ¼ before the curly bracket, shows the probability of finding an electron at this energy level ($E = \varepsilon + \frac{U}{2} + \frac{\alpha}{2} + B$). If we sum up the probabilities of finding electrons in the case of all eight fractions in (13), we get the probability of a reliable event.

From the analysis of formula (13) one could conclude that the energy spectrum of the dimer includes eight energy levels (see, for example, [30 - 34]). However, this is not entirely true, since at some energy levels the electron can be found with a probability equal to zero. To do this, it is necessary to calculate all the thermodynamic averages included in the numerators of the fractions included in formula (13). For this, all the Fourier transforms of the anticommutator Green's functions were successively calculated, which determine the thermodynamic averages we need according to the fluctuation-dissipation theorem, then, using the fluctuation-dissipation theorem, expressions for the thermodynamic averages were obtained, it is clear that other thermodynamic averages also arose in this case, for the determination of which the necessary calculations were also made. Having solved the resulting system of algebraic equations for the thermodynamic averages in the ground state, we obtained the following results in the case of strong correlations:

$$\langle a_{1\uparrow}^+ a_{2\uparrow}\rangle = \langle \hat{n}_{1\uparrow}\hat{n}_{2\uparrow}\rangle = \langle \hat{n}_{1\uparrow} a_{1\downarrow}^+ a_{2\downarrow}\rangle = \langle \hat{n}_{1\uparrow} a_{2\downarrow}^+ a_{1\downarrow}\rangle = \langle a_{2\uparrow}^+ a_{2\downarrow}^+ a_{1\uparrow} a_{2\downarrow}\rangle = \langle \hat{n}_{1\uparrow}\hat{n}_{2\uparrow}\hat{n}_{2\downarrow}\rangle = 0,$$

$$\langle \hat{n}_{1\uparrow}\hat{n}_{1\downarrow}\rangle = \langle \hat{n}_{2\uparrow}\hat{n}_{2\downarrow}\rangle = 0,$$

$$\langle \hat{n}_{1\downarrow}\rangle = \langle \hat{n}_{1\uparrow}\rangle = \langle \hat{n}_{2\downarrow}\rangle = \langle \hat{n}_{2\uparrow}\rangle = \langle \hat{n}_{1\uparrow}\hat{n}_{2\downarrow}\rangle = \langle \hat{n}_{1\downarrow}\hat{n}_{2\uparrow}\rangle = \langle a_{2\uparrow}^+ a_{1\downarrow}^+ a_{1\uparrow} a_{2\downarrow}\rangle = 1/2.$$

Let us note that the probability of finding two electrons with oppositely oriented spin projections at one site in the ground state in the strong correlation regime $\langle \hat{n}_{1\uparrow}\hat{n}_{1\downarrow}\rangle$ is zero, which corresponds to the result of [35], see Fig. 3 in Ref. [35] in the case of strong correlations. If the thermodynamic average corresponding to the hopping of an electron from a site to an adjacent site in the ground state in the strong correlation regime $\langle a_{1\uparrow}^+ a_{2\uparrow}\rangle$ is zero, then the probability of a simultaneous pairwise hopping of one electron, for example, from the first site to another, of another electron from the second site to the first is equal to ½: ($\langle a_{2\uparrow}^+ a_{1\downarrow}^+ a_{1\uparrow} a_{2\downarrow}\rangle = 1/2$). Thus, at large values of the Coulomb potential compared to the hopping integral, it is more advantageous for electrons to perform pairwise hops. Let us add to this that the probability of detecting one electron at one site of the dimer, and the second electron with the oppositely oriented spin projection at another site is also equal to ½: ($\langle \hat{n}_{1\uparrow}\hat{n}_{2\downarrow}\rangle = \langle \hat{n}_{1\downarrow}\hat{n}_{2\uparrow}\rangle = 1/2$). That is, the ground state of the dimer is antiferromagnetic. It can be hypothesized that in the case of a larger number of sites, for example, in carbon nanotubes, fullerenes, graphene, biphenylene, it is energetically more favorable for electrons to move in the ground state in strongly correlated systems in pairs. From this point of view, the work [36] is of interest, in which the Hamiltonian of the Hubbard model for a dimer is diagonalized in the spirit of the BCS theory.

If we substitute the resulting thermodynamic averages into formula (13), we obtain the following expression for the Fourier transform of the anticommutator Green's function:

$$\langle\langle a_{1\uparrow}^+ | a_{1\uparrow}\rangle\rangle_E = \frac{i}{2\pi}\left\{\frac{1/4}{E-\varepsilon-B} + \frac{1/4}{E-\varepsilon+B} + \frac{1/4}{E-\varepsilon-U-B} + \frac{1/4}{E-\varepsilon-U+B}\right\}. \qquad (14)$$

Earlier we noted that the poles of the Fourier transform of the anticommutator Green's function (14) determine the energy spectrum of the quantum system under study. The energy spectrum for the parameter values $U = 7\ eV, B = -1\ eV$ is shown in Fig. 1 (the choice of such values of the dimer parameters is due to the fact that they correspond, on average, to the corresponding parameters in carbon fullerenes, a nanotube; of course, the presence of an essentially triple bond between carbon atoms in a free dimer will lead to an increase in the transfer integral compared to jumps in fullerenes, nanotubes, but we are interested in the dimer as a structural element of the fullerene). But in this case, it was necessary at first to determine the value of the self-energy ε, based on the equation for the chemical potential. A graphical representation of the equation for the chemical potential is shown in Fig. 2. On the ordinate axis, we have the number of electrons $N_e$. In the case where there is an average of one electron per site, we have $N_e = 2$. This value of $N_e$ corresponds to the value $\varepsilon = -\frac{U}{2} = -3.5\ eV$ on the graph. Any value near $-3.5\ eV$ within the step segment could be taken as the value of the self-energy.

Let us return to Fig. 1. The two upper energy levels $E_{1,2} = \varepsilon + U \pm B$ form the upper Hubbard subband (analogous to the conduction band), the other two lower energy levels $E_{3,4} = \varepsilon \pm B$ form the lower Hubbard subband (analogous to the valence electron band). At each of the four energy levels, the π-electron can be found with a probability of 1/4. It should be noted that one of the $\pi$-electrons in the ground state will be at the energy level $\varepsilon + B$, the second $\pi$-electron will be at the energy level $\varepsilon - B$. The degeneracy multiplicities of the states corresponding to energies $\varepsilon \pm B$ are equal to one, so from the point of view of symmetry theory these energy levels could be marked as $a_g$ in the case of the energy level $\varepsilon + B$ (the symbol $a$ indicates that the degeneracy multiplicity is one, the symbol $g$ indicates that the state is even) and as $a_u$ in the case of the energy level $\varepsilon - B$ (the symbol $u$ indicates that the state is odd). The following Fig. 3 shows a graph of the dependence for the density of the electron state, the Van Hove singularities correspond to the poles of the Fourier transform of the Green's function. When modeling the delta function, the half-width C was taken to be 0.1.

**Hubbard model for a dimer in the approximation of static fluctuations**

Since the solution of applied problems in physics and chemistry of strongly correlated systems within the framework of the Hubbard model requires the use of various kinds of approximations, the exact solution of the dimer can be used to develop methods for the approximate solution of the Hubbard model that adequately describe the properties of strongly correlated systems.

We rewrite the Hamiltonian of the Hubbard model for the dimer (1) as follows:

$$\hat{H} = \hat{H}_0 + \hat{V}, \tag{15}$$

$$\hat{H}_0 = \varepsilon \sum_{f=1}^{2}(\hat{n}_{f\uparrow} + \hat{n}_{f\downarrow}) + B \sum_{f \neq f'}(a^+_{f\uparrow}a_{f'\uparrow} + a^+_{f\downarrow}a_{f'\downarrow}), \tag{16}$$

$$\hat{V} = U \sum_{f=1}^{2} \hat{n}_{f\uparrow}\hat{n}_{f\downarrow}. \tag{17}$$

The equations of motion for the π-electron creation operators are:

$$\frac{da_{1\uparrow}^+(\tau)}{d\tau} = \varepsilon a_{1\uparrow}^+(\tau) + B a_{2\uparrow}^+(\tau) + U\hat{n}_{1\downarrow}a_{1\uparrow}^+(\tau), \tag{18}$$

$$\frac{da_{2\uparrow}^+(\tau)}{d\tau} = \varepsilon a_{2\uparrow}^+(\tau) + B a_{1\uparrow}^+(\tau) + U\hat{n}_{2\downarrow}a_{2\uparrow}^+(\tau). \tag{19}$$

We introduce the operator of fluctuation of the number of π-electrons $\Delta\hat{n}_{1\downarrow}(\tau)$ as follows:

$$\hat{n}_{1\downarrow} = \langle\hat{n}_{1\downarrow}\rangle + \Delta\hat{n}_{1\downarrow}. \tag{20}$$

Where $\langle\hat{n}_{1\downarrow}\rangle$ is the thermodynamic average describing the average number of electrons at site 1 with spin projection ↓. We rewrite (18) as:

$$\frac{da_{1\uparrow}^+(\tau)}{d\tau} = \varepsilon' a_{1\uparrow}^+(\tau) + B a_{2\uparrow}^+(\tau) + U\Delta\hat{n}_{1\downarrow}a_{1\uparrow}^+(\tau), \tag{21}$$

Let us introduce the representation of the "interaction representation type" as follows:

$$a_{1\uparrow}^+(\tau) = e^{\hat{H}\tau}a_{1\uparrow}^+(0)e^{-\hat{H}\tau} = e^{\hat{H}_0\tau}e^{-\hat{H}_0\tau}e^{\hat{H}\tau}a_{1\uparrow}^+(0)e^{-\hat{H}\tau}e^{\hat{H}_0\tau}e^{-\hat{H}_0\tau} = e^{\hat{H}_0\tau}\tilde{a}_{1\uparrow}^+(\tau)e^{-\hat{H}_0\tau}. \tag{22}$$

Where $\hat{H}_0$ is the operator in Hamiltonian (16) describing the electron self-energy, taking into account the redefinition $\varepsilon \to \varepsilon + U\langle\hat{n}_{1\downarrow}\rangle$. For the creation operator $\tilde{a}_{1\uparrow}^+(\tau)$, introduced in (22), the equation of motion has the form:

$$\frac{d}{d\tau}\tilde{a}_{1\uparrow}^+(\tau) = U\Delta\tilde{n}_{1\downarrow}\tilde{a}_{1\uparrow}^+(\tau). \tag{23}$$

On the right side of (23), as expected, we obtained an operator of higher order. The operator of particle number fluctuations $\Delta\tilde{n}_{1\downarrow}(\tau)$ at this step is declared to be the operator of static fluctuations: $\Delta\tilde{n}_{1\downarrow}(\tau) \approx \Delta n_{1\downarrow}(0)$. We call this approximation the approximation of static fluctuations. In justification of the use of such an approximation, the following can be noted: the particle number operator in the Heisenberg representation $\Delta\hat{n}_{1\downarrow}(\tau)$ commutes with the "perturbation" Hamiltonian $\hat{V}$: $[\hat{V}, \Delta\hat{n}_{1\downarrow}(\tau)] = [\hat{V}, \hat{n}_{1\downarrow}(\tau)] = 0$, for this reason the equation of motion for the particle number fluctuation operator will have the form:

$$\frac{d}{d\tau}\Delta\hat{n}_{1\downarrow}(\tau) = \frac{d}{d\tau}\hat{n}_{1\downarrow}(\tau) = [\hat{H}, \hat{n}_{1\downarrow}(\tau)] = [\hat{H}_0, \hat{n}_{1\downarrow}(\tau)].$$

Thus, when calculating the commutator $[\hat{H}, \Delta\hat{n}_{1\downarrow}(\tau)]$ we can set $\hat{H} = \hat{H}_0$ – the main time dependence of the particle number fluctuation operator is contained in $\hat{H}_0$. Therefore, we can make the following approximation:

$$\Delta\tilde{n}_{1\downarrow}(\tau) = e^{-\hat{H}_0\tau}e^{\hat{H}\tau}\Delta\hat{n}_{1\downarrow}(0)e^{-\hat{H}\tau}e^{\hat{H}_0\tau} \approx e^{-\hat{H}_0\tau}e^{\hat{H}_0\tau}\Delta\hat{n}_{1\downarrow}(0)e^{-\hat{H}_0\tau}e^{\hat{H}_0\tau} = \Delta\hat{n}_{1\downarrow}(0).$$

In this case, the equation of motion for the operator on the right side of formula (23) is easily closed:

$$\frac{d}{d\tau}\Delta\tilde{n}_{1\downarrow}(\tau)\tilde{a}_{1\uparrow}^+(\tau) = \Delta\hat{n}_{1\downarrow}(0)\frac{d}{d\tau}\tilde{a}_{1\uparrow}^+(\tau) = U\langle\hat{n}_{1\downarrow}\rangle(1-\langle\hat{n}_{1\downarrow}\rangle)\tilde{a}_{1\uparrow}^+(\tau) +$$

$$+U(1-2\langle\hat{n}_{1\downarrow}\rangle)\Delta\hat{n}_{1\downarrow}(0)\tilde{a}_{1\uparrow}^+(\tau). \qquad (24)$$

The system of differential equations (23), (24) for the creation operator has the following solution:

$$\tilde{a}_{1\uparrow}^+(\tau) = \left((1-\langle\hat{n}_{1\downarrow}\rangle)\right)a_{1\uparrow}^+(0)\exp(-U\langle\hat{n}_{1\downarrow}\rangle\tau) + \langle\hat{n}_{1\downarrow}\rangle a_{1\uparrow}^+(0)\exp(U(1-\langle\hat{n}_{1\downarrow}\rangle)\tau)+$$

$$+\Delta\hat{n}_{1\downarrow}(0)a_{1\uparrow}^+(0)\{\exp(U(1-\langle\hat{n}_{1\downarrow}\rangle)\tau) - \exp(-U\langle\hat{n}_{1\downarrow}\rangle\tau)\}, \qquad (25)$$

We will not give the second solution of the system of differential equations, since for further reasoning it is sufficient for us to have solution (25). From (22) it follows that the solution for the creation operator in the Heisenberg representation will have the form:

$$a_{1\uparrow}^+(\tau) = \{(1-\langle\hat{n}_{1\downarrow}\rangle) + \langle\hat{n}_{1\downarrow}\rangle\exp(U\tau)\}\exp(-U\langle\hat{n}_{1\downarrow}\rangle\tau)\,e^{\hat{H}_0\tau}a_{1\uparrow}^+(0)e^{-\hat{H}_0\tau} +$$

$$\{\exp(U\tau) - 1\}\exp(-U\langle\hat{n}_{1\downarrow}\rangle\tau)\,e^{\hat{H}_0\tau}\Delta\hat{n}_{1\downarrow}(0)e^{-\hat{H}_0\tau}e^{\hat{H}_0\tau}a_{1\uparrow}^+(0)e^{-\hat{H}_0\tau}. \qquad (26)$$

The equation of motion for the operator included in (26) $\bar{a}_{1\uparrow}^+(\tau) = e^{\hat{H}_0\tau}a_{1\uparrow}^+(0)e^{-\hat{H}_0\tau}$ has the form:

$$\frac{d}{d\tau}\bar{a}_{1\uparrow}^+(\tau) = \varepsilon'\bar{a}_{1\uparrow}^+(\tau) + B\bar{a}_{2\uparrow}^+(\tau). \qquad (27)$$

In a similar way, we can obtain the equation of motion for the operator $\bar{a}_{2\uparrow}^+(\tau) = e^{\hat{H}_0\tau}a_{2\uparrow}^+(0)e^{-\hat{H}_0\tau}$:

$$\frac{d}{d\tau}\bar{a}_{2\uparrow}^+(\tau) = \varepsilon'\bar{a}_{2\uparrow}^+(\tau) + B\bar{a}_{1\uparrow}^+(\tau). \qquad (28)$$

The solutions of the system of coupled differential equations (27), (28) have the form:

$$\bar{a}_{1\uparrow}^+(\tau) = \frac{1}{2}e^{\varepsilon'\tau}\{a_{1\uparrow}^+(0)[e^{B\tau}+e^{-B\tau}] + a_{2\uparrow}^+(0)[e^{B\tau}-e^{-B\tau}]\}, \qquad (29)$$

$$\bar{a}_{2\uparrow}^+(\tau) = \frac{1}{2}e^{\varepsilon'\tau}\{a_{2\uparrow}^+(0)[e^{B\tau}+e^{-B\tau}] + a_{1\uparrow}^+(0)[e^{B\tau}-e^{-B\tau}]\}. \qquad (30)$$

Substituting expressions (29), (30) into formula (26) we can obtain a solution for the particle creation operator in the form:

$$a_{1\uparrow}^+(\tau) = \frac{1}{2}e^{\varepsilon'\tau}\{(1-\langle\hat{n}_{1\downarrow}\rangle) + \langle\hat{n}_{1\downarrow}\rangle\exp(U\tau)\}\exp(-U\langle\hat{n}_{1\downarrow}\rangle\tau)\cdot\{a_{1\uparrow}^+(0)[e^{B\tau}+e^{-B\tau}]+$$
$$a_{2\uparrow}^+(0)[e^{B\tau}-e^{-B\tau}]\} + \frac{1}{2}e^{\varepsilon'\tau}\{\exp(U\tau)-1\}\exp(-U\langle\hat{n}_{1\downarrow}\rangle\tau)\cdot\{\Delta\bar{n}_{1\downarrow}(\tau)a_{1\uparrow}^+(0)[e^{B\tau}+e^{-B\tau}]+$$
$$\Delta\bar{n}_{1\downarrow}(\tau)a_{2\uparrow}^+(0)[e^{B\tau}-e^{-B\tau}]\}. \tag{31}$$

Let us note that when calculating the anticommutator Green's function, we obtain the thermodynamic average $\langle\Delta\bar{n}_{1\downarrow}(\tau)\rangle$ according to formula (31). Having written down the equations of motion for the operators included in the fluctuation operator $\Delta\bar{n}_{1\downarrow}(\tau)$ and having solved these differential equations, we obtain the solutions:

$$\bar{a}_{1\downarrow}^+(\tau) = (a_{1\downarrow}^+(0)\cosh(B\tau) + a_{2\downarrow}^+(0)\sinh(B\tau))\exp(\varepsilon'\tau),$$

$$\bar{a}_{1\downarrow}(\tau) = (a_{1\downarrow}(0)\cosh(B\tau) - a_{2\downarrow}(0)\sinh(B\tau))\exp(\varepsilon'\tau).$$

Using the obtained solutions, we obtain the following expression for $\Delta\bar{n}_{1\downarrow}(\tau)$:

$$\Delta\bar{n}_{1\downarrow}(\tau) = \hat{n}_{1\downarrow}(0)\cosh^2(B\tau) - \langle\hat{n}_{1\downarrow}\rangle - \hat{n}_{1\downarrow}(0)\sinh^2(B\tau)$$
$$+ \left(a_{2\downarrow}^+(0)a_{1\downarrow}(0) - a_{1\downarrow}^+(0)a_{2\downarrow}(0)\right)\sinh(B\tau)\cosh(B\tau).$$

Since $\langle\hat{n}_{1\downarrow}\rangle = \langle\hat{n}_{2\downarrow}\rangle = 1/2$, and also $\langle a_{2\downarrow}^+ a_{1\downarrow}\rangle = \langle a_{1\downarrow}^+ a_{2\downarrow}\rangle$, we get that $\langle\Delta\bar{n}_{1\downarrow}(\tau)\rangle = 0$.

Solution (31) characterizes the evolution of a quantum system described by Hamiltonian (15) in the approximation of static fluctuations. Using (31), we calculate the Fourier transform of the anticommutator Green's function:

$$\langle\langle a_{1\uparrow}^+|a_{1\uparrow}\rangle\rangle_E = \frac{i}{2\pi}\frac{1}{2}\left\{\frac{1-\langle\hat{n}_{1\downarrow}\rangle}{E-\varepsilon-B} + \frac{1-\langle\hat{n}_{1\downarrow}\rangle}{E-\varepsilon+B} + \frac{\langle\hat{n}_{1\downarrow}\rangle}{E-\varepsilon-U-B} + \frac{\langle\hat{n}_{1\downarrow}\rangle}{E-\varepsilon-U+B}\right\}. \tag{32}$$

Based on expression (32) and using the fluctuation-dissipation theorem, it is easy to obtain that in the case of two electrons per two sites (in the case of a band that is exactly half filled) $\langle\hat{n}_{1\downarrow}\rangle = \langle\hat{n}_{1\uparrow}\rangle = \frac{1}{2}$, substituting $\langle\hat{n}_{1\downarrow}\rangle = \frac{1}{2}$ into (32) we obtain the final expression for the Fourier transform of the anticommutator Green's function:

$$\langle\langle a_{1\uparrow}^+|a_{1\uparrow}\rangle\rangle_E = \frac{i}{2\pi}\left\{\frac{1/4}{E-\varepsilon-B} + \frac{1/4}{E-\varepsilon+B} + \frac{1/4}{E-\varepsilon-U-B} + \frac{1/4}{E-\varepsilon-U+B}\right\}. \tag{33}$$

Thus, the Fourier transform of the anticommutator Green's function in the approximation of static fluctuations (33) in the case of a dimer coincided with the exact solution (14). This gives hope that the application of the approximation of static fluctuations in the study of strongly correlated systems will adequately describe the physicochemical properties of these systems. The approximation of static fluctuations was developed and used in works [37- 43]. The application of the static fluctuation approximation made it possible to theoretically explain the electronic structure, optical absorption spectra of carbon and gold fullerenes, carbon and gold nanotubes, and a biphenylene network.

Let us now calculate the ground state energy, ionization energy, and electron affinity in the case of a dimer. The ground state energy (the average value of the dimer energy at temperature $T \to 0\ K$) is equal to:

$$E_0 = \langle \hat{H} \rangle = 4\varepsilon \langle \hat{n}_{1\uparrow} \rangle + 4 B \langle a_{1\uparrow}^+ a_{2\uparrow} \rangle + 2 U \langle \hat{n}_{1\uparrow} \hat{n}_{1\downarrow} \rangle. \tag{34}$$

Above we defined the thermodynamic averages included in formula (34) in the ground state in the case of strong correlations. Since the ionization energy and electron affinity in the case of fullerenes are determined at sufficiently high temperatures (for example, at 777 K in [44], at 873 K in [45], at 932 K in [46]), the average value of the dimer energy and thermodynamic averages must be calculated at finite temperature values. Knowing the solution for the particle creation operator at the first site, we can obtain the following formula for the thermodynamic average describing the electron transfer from the second site to the first site:

$$\langle a_{1\uparrow}^+ a_{2\uparrow} \rangle = \frac{1}{2} \left( \frac{(1 - \langle \hat{n}_{1\downarrow} \rangle)}{e^{\beta(\varepsilon+B)} + 1} - \frac{(1 - \langle \hat{n}_{1\downarrow} \rangle)}{e^{\beta(\varepsilon-B)} + 1} + \frac{\langle \hat{n}_{1\downarrow} \rangle}{e^{\beta(\varepsilon+U+B)} + 1} - \frac{\langle \hat{n}_{1\downarrow} \rangle}{e^{\beta(\varepsilon+U-B)} + 1} \right). \tag{35}$$

Having formula (35), we can construct graphs of the dependence of thermodynamic averages for three cases: 1) for an electrically charged ion $C_2^+$, setting $\langle \hat{n}_{1\downarrow} \rangle = 1/4$, choosing any value for the self-energy of the π-electron $\varepsilon$ within the interval $-1 < \varepsilon < 1$, based on the graphical representation of the equation for the chemical potential shown in Fig. 2; 2) for a neutral dimer $C_2$, setting $\langle \hat{n}_{1\downarrow} \rangle = 2/4$, choosing any value for the self-energy of the π-electron $\varepsilon$ within the interval $-6 < \varepsilon < -1$; 3) for an electrically charged ion $C_2^-$, putting $\langle \hat{n}_{1\downarrow} \rangle = 3/4$, choosing any value for the self-energy of the π-electron $\varepsilon$ within the interval $-8 < \varepsilon < -6$. These graphs are shown in Fig. 4. This figure shows the correlation function $\langle a_{1\uparrow}^+ a_{2\uparrow} \rangle$ depending on the ratio of the Coulomb potential $U$ to the transfer integral $B$. In our case, the parameters of the Hubbard model are $U = 7\ eV, B = -1\ eV$, then the ratio of the Coulomb potential $U$ to the transfer integral $B$, taking into account the minus sign in front of the ratio, is 7. Therefore, we will be most interested in the region when this ratio is greater than six, when the regime of strong electron correlations is realized. In the case of a neutral dimer $C_2$, the probability of an electron jumping from site to site in this region is very small due to the fact that the Coulomb potential $U$ significantly exceeds the transfer integral; a π-electron is practically unable to overcome the potential barrier in order to end up at the neighboring site. The behavior of curve 1 for the case of a positively charged ion $C_2^+$ is quite understandable, since there will always be no electron at one of the dimer sites, which means that there will be virtually no potential barrier on the path of a π-electron moving to this site. The behavior of curve 3, corresponding to the negatively charged ion $C_2^-$, in this region can be explained as follows. On average, there is one π-electron at each site in this case, and there is also an "extra" third π-electron. The behavior of curve 2 shows that the transition of a π-electron from a site with one π-electron to a neighboring site with one π-electron is essentially impossible. There remains the case when the transition in accordance with the Pauli principle occurs from a site where there are two π-electrons to a neighboring site with one π-electron. In this case, the π-electron moving to the neighboring site is accelerated by the Coulomb field of another π-electron at this site, as a result of which the electron energy will be significantly higher, so the π-electron overcomes the Coulomb barrier relatively easily.

Let us calculate the thermodynamic average $\langle \hat{n}_{1\uparrow} \hat{n}_{1\downarrow} \rangle$, which describes the probability of finding two π-electrons with oppositely oriented spin projections at one site. The formula for it is:

$$\langle \hat{n}_{1\uparrow} \hat{n}_{1\downarrow} \rangle = \frac{1}{2} \left( \frac{\langle \hat{n}_{1\downarrow} \rangle}{e^{\beta(\varepsilon+U+B)} + 1} - \frac{\langle \hat{n}_{1\downarrow} \rangle}{e^{\beta(\varepsilon+U-B)} + 1} \right). \tag{36}$$

The dependence of the correlation function $\langle \hat{n}_{1\uparrow} \hat{n}_{1\downarrow} \rangle$ on the ratio of the Coulomb potential to the transfer integral in the region of strong correlations shown in Fig. 5 is quite logical – the more electrons the C2 dimer contains, the more likely it is that two π-electrons can be found at one site of the dimer.

Fig. 6 shows the ground state energy for the electrically neutral dimer $C_2$ (1), the ground state energy for the positively charged ion $C_2^+$ (2), and the ground state energy for the negatively charged ion $C_2^-$ (3) depending on the ratio $x = -U/B$ near the parameter values $U = 7\ eV, B = -1\ eV$. Having determined the difference between the energies for the case when the dimer contains one electron (the dimer ion $C_2^+$) and for the case when the dimer contains two electrons with oppositely oriented spin projections (the neutral dimer $C_2$) at $x = 7$, we can obtain that the ionization energy of the dimer is $11.87\ eV$. This ionization energy value correlates with the experimental values of the ionization energy of the carbon molecule $C_2$ $E_I = 11.79\ eV$ in [47] and $E_I = 11.87\ eV$ in [48]. Having determined in Fig. 6 difference between the energies of the ground state for the neutral dimer and for the dimer ion $C_2^-$ containing three electrons, we can calculate the electron affinity energy $A_E$ for the dimer - $A_E = 3.39\ eV$. This value for the electron affinity energy is consistent with the electron affinity energy in the case of the carbon molecule $C_2$ in [49, 50] $A_E = 3.391 \pm 0{,}017\ eV$.

**Conclusions**

Thus, in the present work both the exact solution of the dimer as a structural element of carbon fullerene, nanotubes, graphene, biphenylene, and the approximate solution in the "approximation of static fluctuations" are produced. Since the Fourier transforms of the anticommutator Green's functions for the exact solution and the approximate solution coincided, it can be concluded that the approximation of static fluctuations describes the properties of strongly correlated systems quite adequately. The calculated values of the ionization energy of the dimer and the electron affinity of the dimer correlate with the experimental values of these quantities for the carbon molecule $C_2$.

Notes and references.

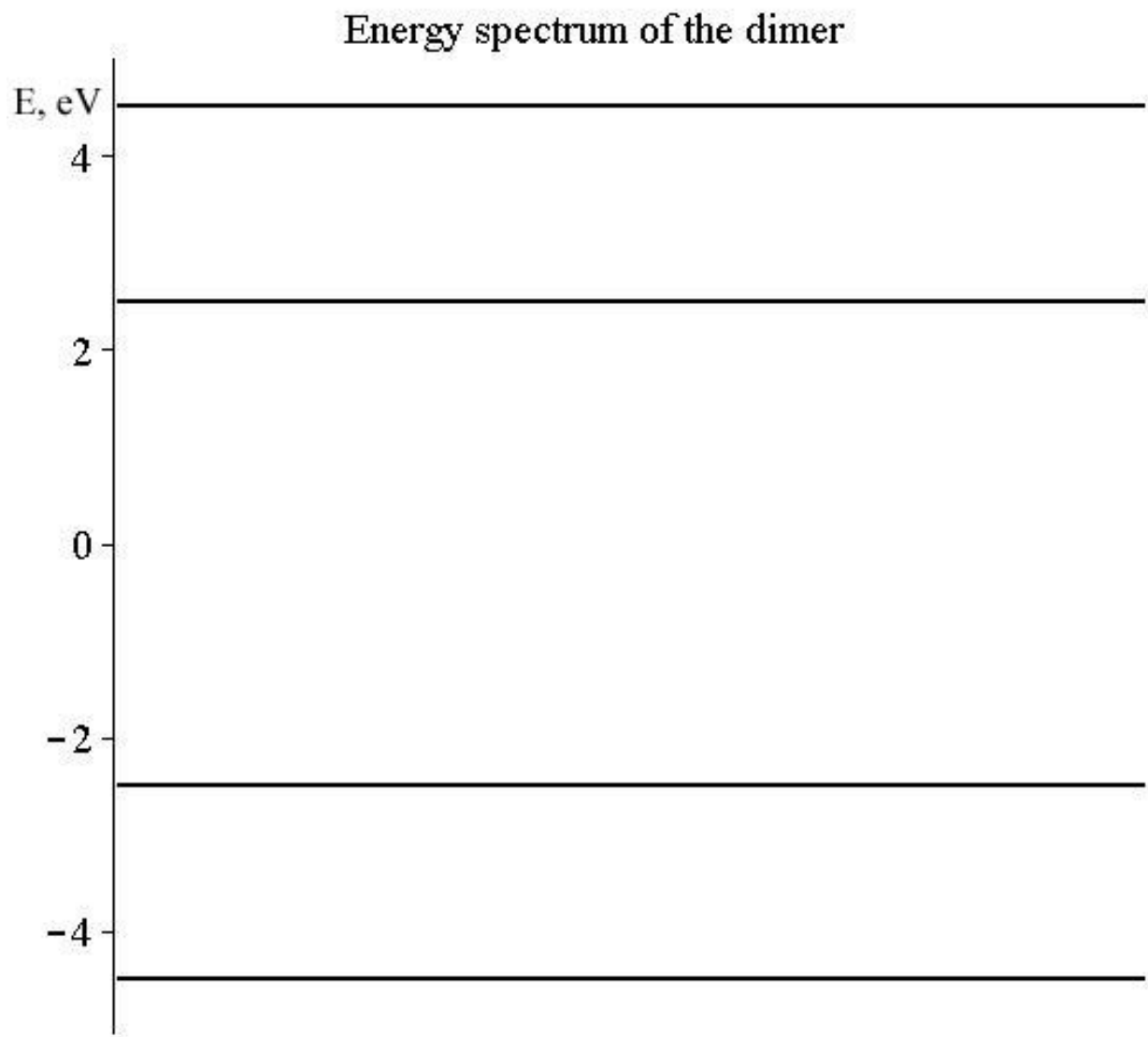

Fig. 1. Energy spectrum for parameter values $U = 7\ eV, B = -1\ eV, \varepsilon = -3.5\ eV$.

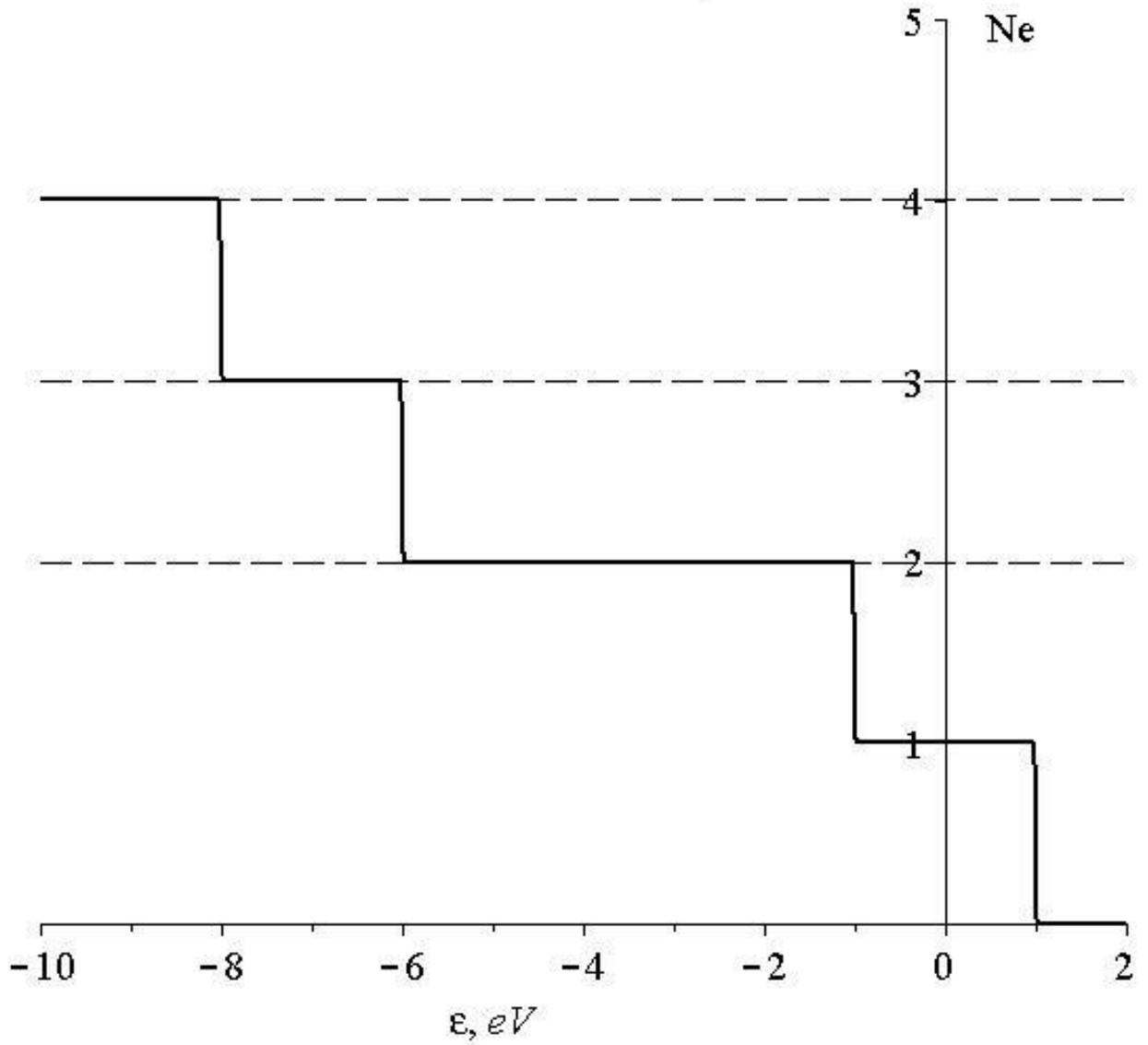

Fig. 2. Graphical representation of the equation for the chemical potential with the parameter values $U = 7\ eV, B = -1\ eV$.

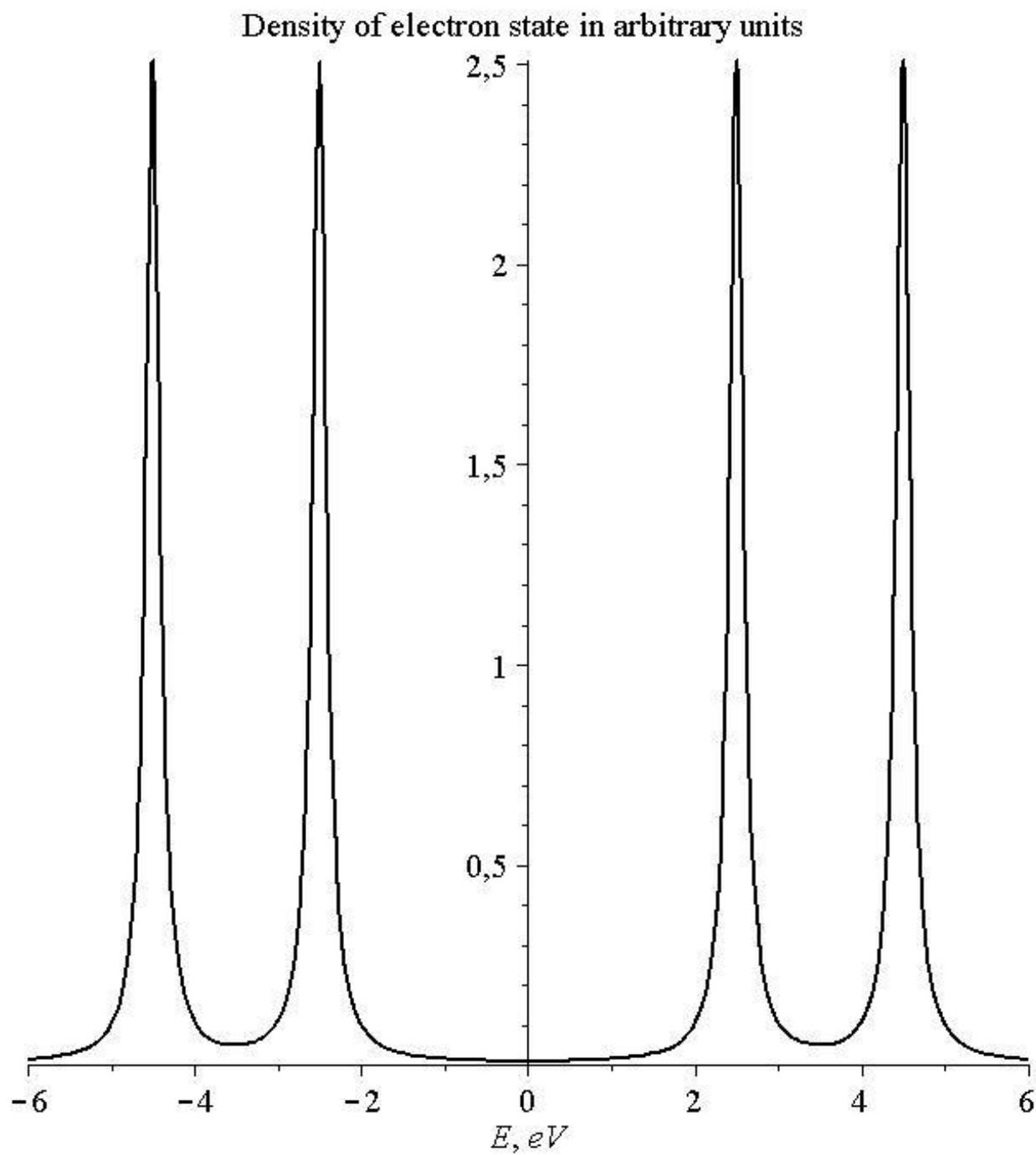

Fig. 3. Density of the electron state at the parameter values $U = 7\ eV, B = -1\ eV, \varepsilon = -3.5\ eV, C = 0.1$.

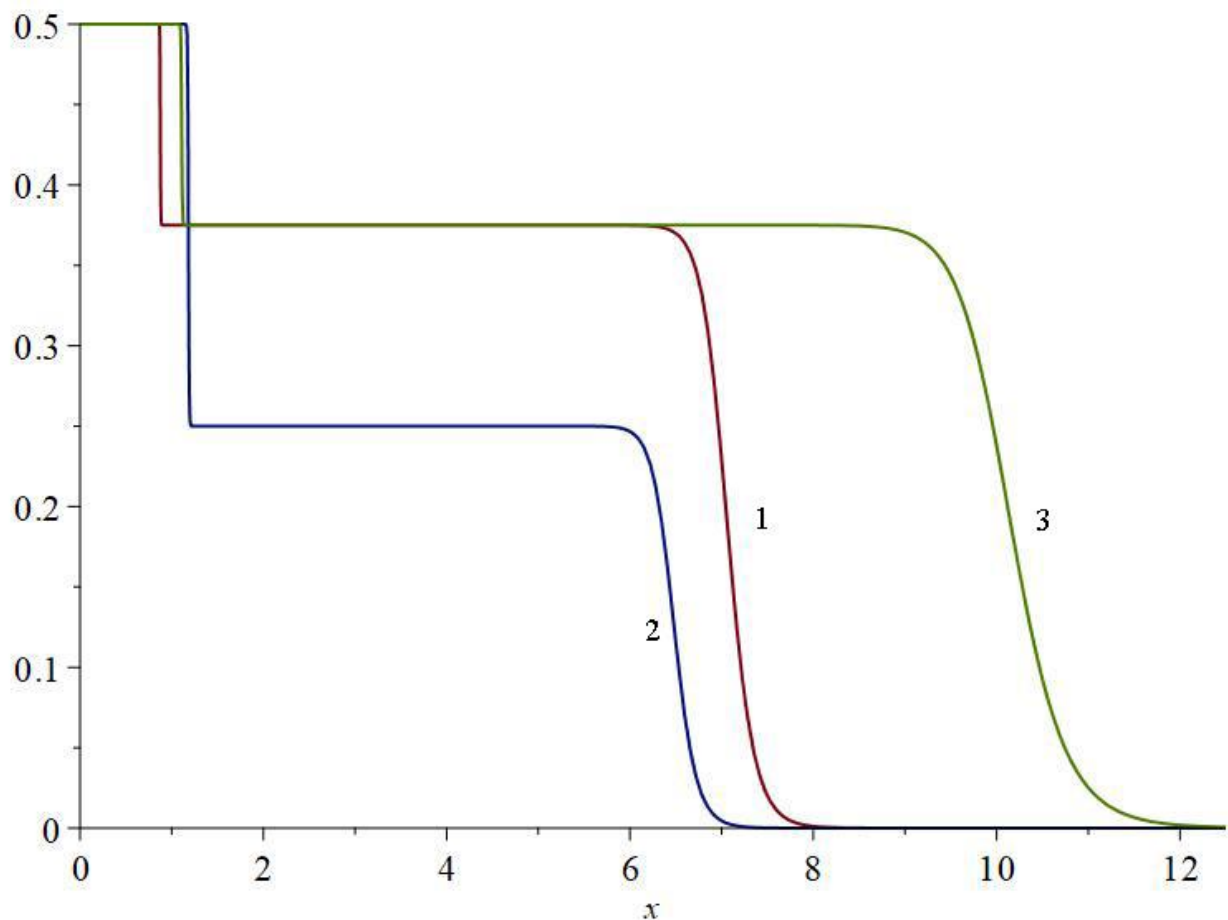

Fig. 4. The correlation function $\langle a_{1\uparrow}^+ a_{2\uparrow}\rangle$, describing the probability of a $\pi$ -electron transition from site 2 to the neighboring site 1 of the dimer depending on the ratio $x = -U/B$. 1 – in the case of a positively charged dimer ion $C_2^+$, 2 – in the case of a neutral dimer $C_2$,, 3 – in the case of a $C_2^-$ - ion.

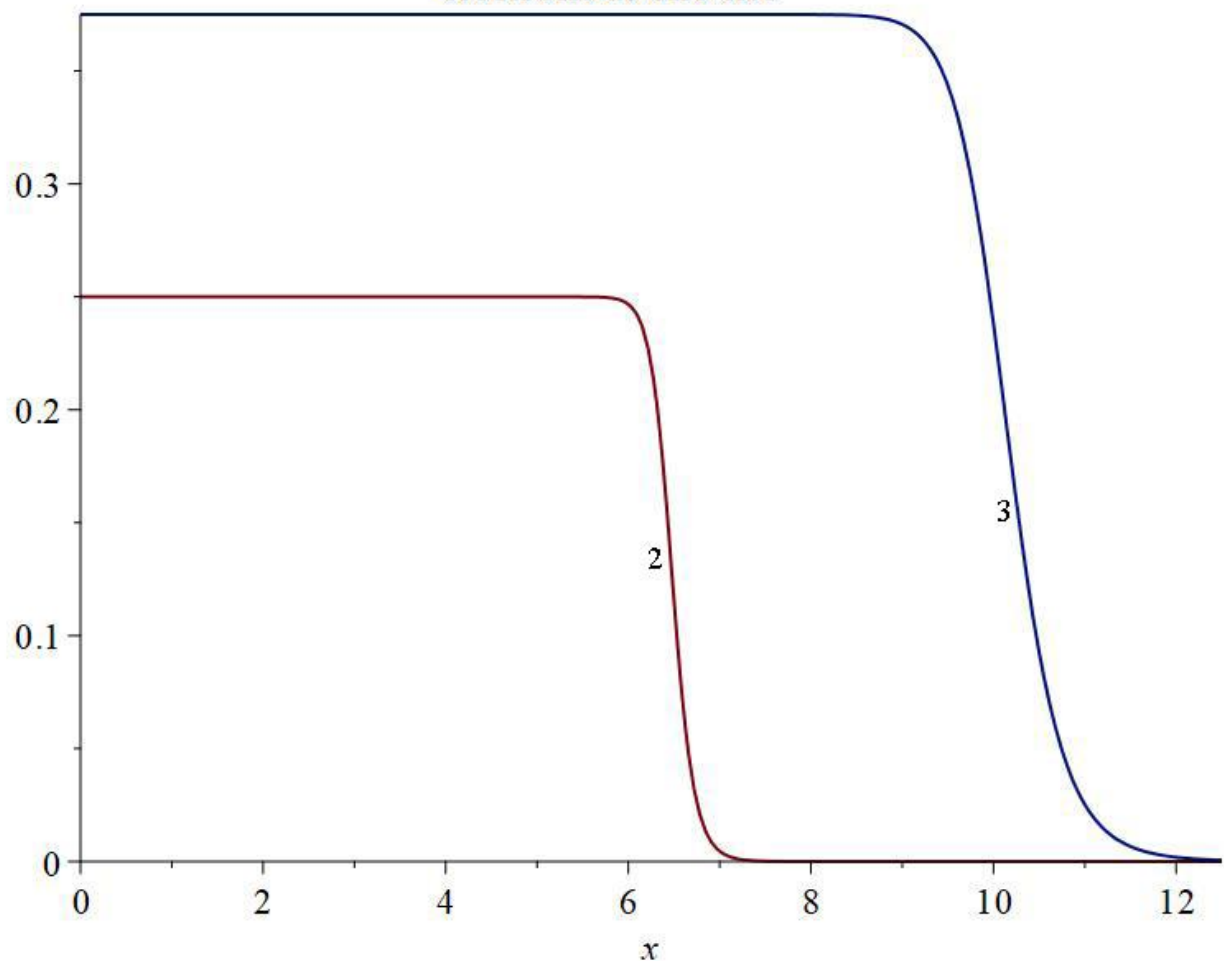

Fig. 5. Correlation function $\langle \hat{n}_{1\uparrow} \hat{n}_{1\downarrow} \rangle$, characterizing the probability of finding two π-electrons with opposite values of spin projections on one site depending on the ratio $x = -U/B$. 2 – in the case of the neutral dimer $C_2$, 3 – in the case of the ion $C_2^-$.

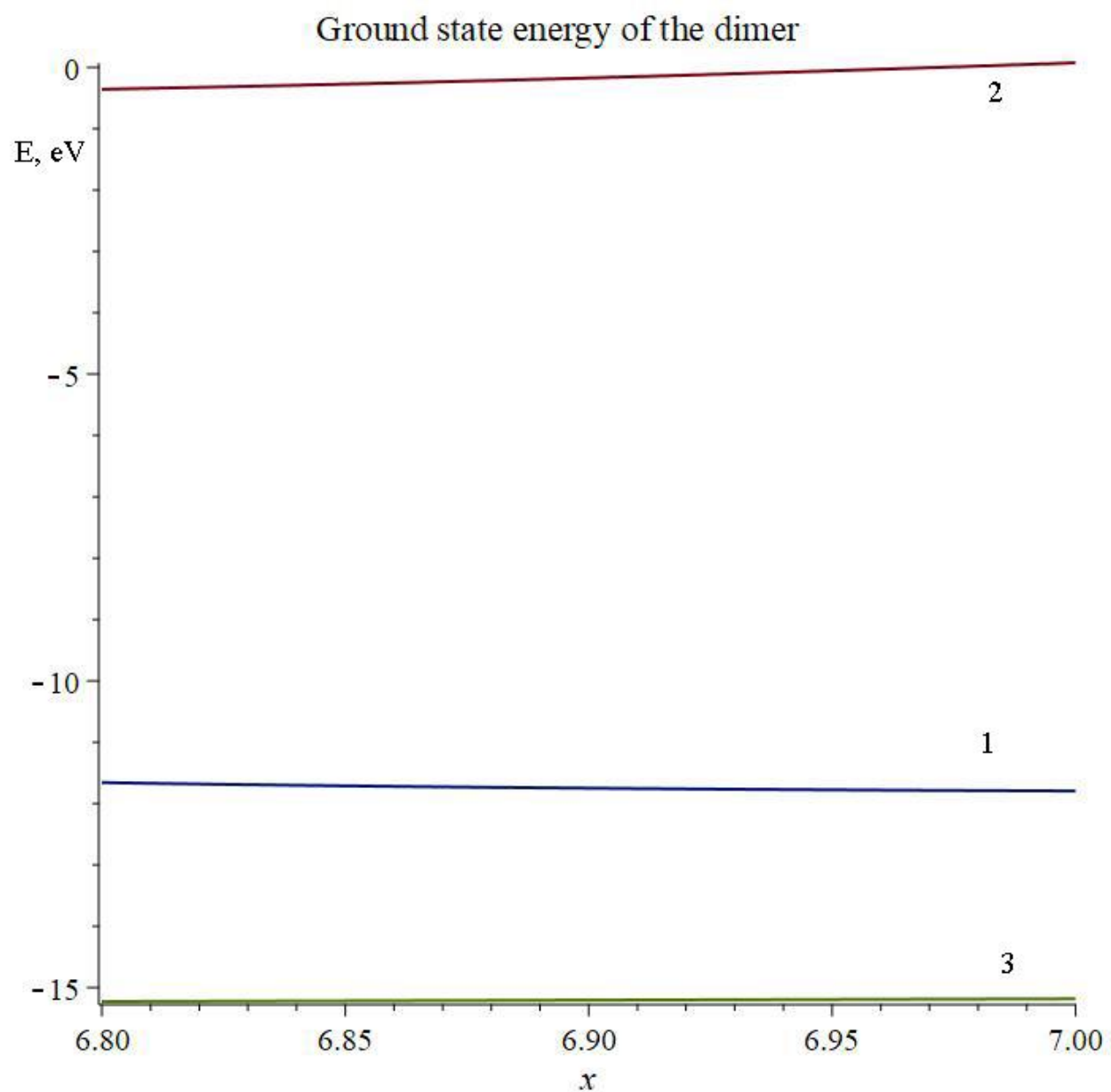

Fig. 6. Ground state energies for the electrically neutral dimer $C_2$ (1), the positively charged ion $C_2^+$ (2) and the negatively charged ion $C_2^-$ (3) depending on the ratio $x = -U/B$.